\documentclass[conference,letter]{IEEEtran}

\usepackage{graphicx} 
\graphicspath{{./figures/}}
\usepackage{float}
\usepackage{array} 
\usepackage{paralist} 
\usepackage{subfig}
\usepackage{fancyhdr} 
\usepackage{amsmath, amsfonts}
\usepackage{mathtools}
\usepackage[most]{tcolorbox}

\newcommand{\ket}[1]{\ensuremath{\left|#1\right\rangle}} 
\newcommand{\bra}[1]{\ensuremath{\left\langle#1\right|}} 

\begin{document}
\title{Quantum Spectral Clustering through a Biased Phase Estimation Algorithm} 
\author{\IEEEauthorblockN{Ammar~Daskin}
}
\maketitle
\begin{abstract}

In this brief paper, we go through the theoretical steps of the spectral clustering on quantum computers by employing the phase estimation and the amplitude amplification algorithms. 
We discuss circuit designs for each step and show how to obtain the clustering solution from the output state.
In addition, we introduce a biased version of the phase estimation algorithm which significantly speeds up the amplitude amplification process. 
The complexity of the whole process is analyzed: it is shown that when the circuit representation of  a data matrix of order $N$ is produced through an ancilla based circuit in which the matrix is written as a sum of $L$ number of Householder matrices; 
the computational complexity is bounded by  $O(2^mLN)$ number of quantum gates. 
Here, $m$ represents the number of qubits (e.g., 6) involved in the phase register of the phase estimation algorithm.
 \end{abstract}
 \begin{IEEEkeywords}
Quantum Cybernetics, Spectral Clustering, Quantum Algorithms.
\end{IEEEkeywords}

 \IEEEpeerreviewmaketitle
\section{Introduction}
In recent years, the research in quantum algorithms for machine learning problems has gained substantial momentum. 
Some of these algorithms include the application of quantum random walk \cite{Kempe2003QRW} to the community detection in quantum networks\cite{Faccin2014community}, quantum nearest neighbor methods \cite{Wiebe2015} for clustering problems, the deep learning in the context of quantum computing \cite{Wiebe2016}, and an accelerated unsupervised learning algorithm with the help of  quantum based subroutines \cite{Aimeur2013quantum}.
Furthermore, quantum algorithms for topological and geometric analysis of data \cite{Lloyd2016quantum} and quantum principal component analysis \cite{Lloyd2014QPCA} are introduced. The computational complexities of these algorithms are exponentially less than the classical algorithms when the data is accessible on a quantum RAM.  
For a broader review of the area we recommend the recent review articles  
and introductory papers such as 
Ref.\cite{Biamonte2016quantum, Arunachalam2017survey} and 
Ref.\cite{Schuld2015introduction}.

 Cluster analysis \cite{Kaufman2009finding} is dividing  a given data points, objects, into clusters in a way that similarities among cluster-members are maximized while inter-cluster similarities are minimized. 
Clustering in machine learning and other fields are done through different approaches. 
One of the best known approach is the centroid based clustering which is also part of spectral clustering algorithms. 
The spectral clustering algorithms define how to obtain a solution for the clustering by using the principal eigenvectors of a provided data matrix or a Laplacian matrix.
Quantum phase estimation is an efficient algorithm to solve eigen-value related problems on quantum computers. In earlier two works \cite{DaskinQPCA2016,Daskin2016qann}, we have showed respectively how to use quantum  phase estimation algorithm to do principal component analysis of a classical data and how this approach can be employed for neural networks using Widrow-Hoff Learning rule.
 In this paper, we employ the same algorithm with slight modifications for clustering. The particular contributions are as follows:
\begin{itemize}
\item for the first time, the spectral clustering is  formulated on quantum computers by introducing a biased quantum phase estimation depicted as a circuit in Fig.\ref{figBPEA}. 
\item The computational complexities of the constituents of the circuit is analyzed individually. Then, the complexity bounds are derived for different cases of the data matrix. The found quantum complexities are discussed and compared with the classical ones.
\end{itemize}

The remaining part of this paper is organized as follows: in the following subsections; for unfamiliar readers,  the k-means clustering and the spectral clustering on classical computers are briefly reviewed.
 Then, in the next section, the principal component analysis introduced in Ref.\cite{DaskinQPCA2016} are described, and it is shown how this method can be used to do spectral clustering on quantum computers.  In the last part, a biased phase estimation algorithm is introduced to decrease the number of iterations in the amplitude amplification process.
Finally, the complexity of the whole process is analyzed and the results are discussed.

\subsection{k-means Clustering and Algorithm}
Spectral clustering algorithms are generally based on obtaining a clustering solution from the eigenvectors of a matrix which represents some form of a given data.
Since some of the spectral algorithms  also involves the k-means algorithm, 
here we first give the description of this algorithm and then describe the spectral clustering.
Given a set of $n$ data vectors, $\textbf{v}_1, \textbf{v}_2,\dots, \textbf{v}_n$,
 $k$-means clustering\cite{Macqueen1967some} tries to find best $k$ centroids for assumed $k$ number of clusters, $S_1 \dots S_k$, by minimizing the following objective function \cite{Dhillon2004unified}:
\begin{equation}
min(\sum_{c=1}^k\sum_{\textbf{v}_i\in S_c} ||\textbf{v}_i-\textbf{m}_c||^2),
\end{equation}
where $\textbf{m}_c$ represents the center of the cluster $S_c$. And $||\textbf{v}_i-\textbf{m}_c||^2$ is the Euclidean distance measure between the data point $\textbf{v}_i$ and the center $\textbf{m}_c$.
The optimization problem defined by the above  objective function is an NP-hard problem; nonetheless, it can be approximately minimized by using $k$-means algorithm, also known as Lloyd's algorithm (which does not necessarily find an optimal solution)\cite{lloyd1982least}. The steps of this algorithm are as follows:
  \begin{enumerate}
  \item Initialize centroids for the clusters.
  \item Assign each data point to the cluster with the closest centroid.
  \item Assign the means of data in  clusters as the new means: i.e., $ \textbf{m}_c = \sum_{\textbf{v}_i\in S_c} 
  \frac{\textbf{v}_i}{|S_c|}$.
  \item Repeat step 2 and 3 until there is no change in the means.
  \end{enumerate}
The quantum version of this algorithm also has been introduced in Ref.\cite{Wiebe2015} and used for the nearest-neighbor classification.
Moreover, in a prior work \cite{Aimeur2007quantum}, quantum subroutines based on Grover's algorithm \cite{Grover1996fast} are presented to speed-up the classical clustering algorithm.  

\subsection{Spectral Clustering}
In this subsection, we mainly follow the related sections of Ref.\cite{Von2007tutorial} and try to briefly summarize the concept of the spectral clustering.
Similarities between data points are most commonly represented by similarity graphs: i.e., undirected weighted graphs in which the vertices $v_i$ and $v_j$ are connected if the data points, $x_i$ and $x_j$ represented by these vertices are similar. And the weights on the edges  $w_{ij}$ indicates the amount of the similarity, $s_{ij}$ between $x_i$ and $x_j$.

The construction of graph $G(V, E)$ from a given data set $\{x_1\dots x_N\}$ with pairwise similarities $s_{ij}$ or distances $d_{ij}$ can be done in many different ways. Three of the famous ones are:
\textbf{The undirected $\epsilon$-neighborhood graph:}
The vertices $v_i$ and $v_j$ are connected if the pairwise distance $d_{ij}$ for $x_i$ and $x_j$ is greater than some threshold $\epsilon$. 
\textbf{The $k$-nearest neighborhood graph:} 
The vertices $v_i$ and $v_j$ are connected if vertex $v_i$ is one of the $k$-nearest neighbor of $v_j$ and vice versa. 
Note that there are some other definitions in the literature even though this is the most commonly used criteria to construct the $k$-nearest neighborhood graph as an undirected graph.
\textbf{The fully connected graph:} This describes a fully connected weighted graph where the weights, $w_{ij}$s, are determined from a similarity function $s_{ij}$: e.g., Gaussian similarity function 
$s_{ij} = exp(-\frac{||x_i-x_j||}{2\sigma^2})$, where $\sigma$ is a control parameter. 

The clustering problem can be described as finding a partition of the graph such that the sum of the weights on the edges from one group to another is very small. And the weights on the edges between vertices inside the same group is very high.
\subsubsection{Similarity Matrix and Its Laplacian}
Let $W$ be the adjacency matrix with the matrix elements, $w_{ij}$ representing the weight on the edge between vertex $v_i$ and vertex $v_j$. 
The eigenvectors of a Laplacian matrix generated from $W$ is the main instrument for the spectral clustering. 
The unnormalized Laplacian for the graph given by $W$ is defined as:
\begin{equation}
L = D-W,
\end{equation}
where $D$ is the diagonal degree matrix with diagonal elements $d_{ii}=\sum_{j=1}^N w_{ij}$.
The matrix $L$ is a symmetric matrix and generally normalized as:
\begin{equation}
\widetilde{L} = D^{-\frac{1}{2}} L D^{-\frac{1}{2}} = I-D^{-\frac{1}{2}} W D^{-\frac{1}{2}}, 
\end{equation}
which preserves the symmetry. $\widetilde{L}$  and $L$ are  semi-positive definite matrices: i.e., their eigenvalues are greater or equal to 0.
 The smallest eigenvalues of both matrices are 0 
 and the elements of the associated eigenvector  are equal to one. 
For undirected graphs with non-negative weights, the multiplicity $k$ of  eigenvalue 0 gives  the number of connected components in the graph.
Clustering is generally done through the first $k$-eigenvectors associated to the smallest eigenvalues of the Laplacian matrix. 
Here, $k$ represents the number of clusters to be constructed from the graph.  
This number describes the smallest eigenvalues $\lambda_1 \dots \lambda_k$ such that  $\gamma_k = |\lambda_k-\lambda_{k+1}|$ gives the largest eigengap among all possible eigengaps.

The clusters are obtained by applying $k$-means algorithm  to the rows of  the matrix $V$ formed by using $k$ column eigenvectors \cite{ShiMalik2000}. The same clustering methodology can also be used after the rows of $V$ are normalized into unit vectors\cite{NgJordanWeiss2002}.

$k$-means clustering can also be done in the subspace of   principal components of  $XX^T$, where $X$ is the centered data matrix\cite{Zha2001spectral,Ding2004k}. 
Furthermore,
for  a given kernel matrix $K$ and a diagonal weight matrix $W$ with $K=WAW$, where $A$ is the adjacency (or similarity) matrix and W is the diagonal weight matrix; the leading $k$ eigenvectors of $W^{1/2}KW^{1/2}$ can be used  for graph partitioning (clustering) by minimizing $||VV^T - YY^T||=2k-2trace(Y^TVV^TY)$ or maximizing $trace(Y^TVV^TY)$. 
Here, $V$ represents the $k$ eigenvectors and $Y$ is the orthonormal indicator matrix representing clusters: $Y=[\mathbf{y_1}, \mathbf{y_2}, \dots, \mathbf{y_k}]$ with 
\begin{equation}
\mathbf{y_j} = \frac{1}{\sqrt{N_j}}[0, \dots, 0, \underbrace{1, \dots, 1,}_{N_j} 0, \dots, 0]^T
\label{Eqy}
\end{equation}
Therefore, the clustering problem turns into the maximization of the following:
\begin{equation}
trace(Y^TVV^TY) = \sum_{j = 1}^k \mathbf{y_j}^TVV^T\mathbf{y_j}.
\label{EqSimilarity}
\end{equation}
 After finding the $k$ eigenvectors, the maximization is done  by running the weighted $k$ means on the rows of $W^{1/2}V$ \cite{Jordan2004learning,Dhillon2004unified}.

\section{Quantum Spectral Clustering}
In the quantum version of the spectral clustering, we will use
 $\bra{\bf{y}}VV^T\ket{\bf{y}}$ as the similarity measure for the clustering. This gives a similar measure to the one in Eq.\eqref{EqSimilarity}. 
 Here $V$ represents the eiegenvectors associated with the non-zero eigenvalues of either a Laplacian $L$, or a data matrix $XX^T$,  or a weighted kernel matrix  $W^{1/2}KW^{1/2}$; and $\ket{\bf{y}}$ represents a basis vector chosen from some arbitrary indicator matrix.
 In the following section, we first describe how to obtain $VV^T\ket{\bf{y}}$ through the phase estimation algorithm then measure the output in the basis $Y$ consisting of $\ket{\bf{y}}$.

\subsection{Principal Component Analysis through Phase Estimation Algorithm}

The phase estimation algorithm \cite{Kitaev,Nielsen2010quantum} finds the phase of the eigenvalue of a unitary matrix. 
If the unitary matrix corresponds to the exponential, $U = e^{iH}$, of a matrix  $H\in \mathcal{R}^{\otimes n}$ which is either a Laplacian matrix or equal to $XX^T$, then the obtained phase represents the eigenvalue of $H$.
Furthermore, while the phase register in the algorithm holds the phase value, the system register holds the associated eigenvector in the output.
When given an arbitrary input, the phase estimation algorithm yields an output state which represents a superposition of the eigenvalues with the probabilities determined from the overlaps of the input and the eigenvectors.
  In Ref.\cite{DaskinQPCA2016}, we have introduced a quantum version of the principal component analysis and showed that one can obtain the eigenvalues in certain desired region and the associated eigenvectors  by applying amplitude amplification algorithm \cite{Mosca1998quantum,Brassard2002,Nielsen2010quantum} to the end of the phase estimation algorithm.
 As done in Ref.\cite{Daskin2016qann} for Widrow-Hoff learning rule,  
 one can eliminate the extra dimensions (zero eigenvalues and eigenvectors) and generate the output state equivalent to $VV^T\ket{\mathbf{y}}$ for some input vector $\ket{\mathbf{y}}$ by using 
  the following amplification iteration:
  \begin{equation}
  Q = U_{PEA}U_s U_{PEA} U_{f_2}. 
  \end{equation}
Here, $U_{PEA}$ represents the phase estimation algorithm applied to $H$ and $U_{f_2}$ and $U_s$ are the marking and the amplification operators described below:
\begin{equation}
U_{f_2} = \left( I^{\otimes m} - 2\ket{\bf{f_2}}\bra{\bf{f_2}}\right) \otimes I^{\otimes{n}}
\end{equation} 
with $\bra{\bf{f_2}} = \frac{1}{\sqrt{N-1}}[0, 1, \dots, 1]$ and $m$ represents the number of qubits in the phase register.
And, 
\begin{equation}
U_{s} = \left( I^{\otimes m+n} - 2\ket{\bf{0}}\bra{\bf{0}}\right). 
\end{equation} 
In the output of the amplitude amplification process, when the first register is in the equal superposition state, the second register holds $VV^T\ket{\mathbf{y}}$.
\subsubsection{With a Biased Phased Estimation Algorithm}
\begin{figure}
\centering
\includegraphics[width = 3in]{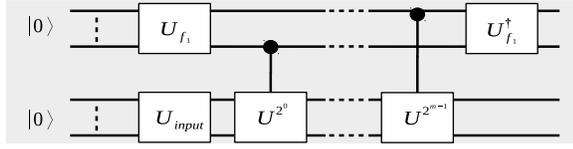}
\caption{Biased Phase Estimation Algorithm \label{figBPEA}}
\end{figure}
 In the output of the phase estimation, the probability values for different eigenpairs are determined from the overlaps of the eigenvectors with the initial vector $\ket{\bf{y}}$: i.e. the inner product of the eigenvector and the initial vector.
Our aim in the amplitude amplification is to eliminate eigenvectors corresponding to the zero eigenvalues. 
The  probability of the chosen states (the eigenvectors corresponding to the non-zero eigenvalues) during the amplitude amplification process oscillates: i.e., when the iteration operator is applied many times, depending on the iterations, the probabilities goes up and down with some frequency.
To speed up the amplitude amplification, to decrease the frequency, we will use a biased phase estimation algorithm which generates the same output but requires less number of iterations:
 In place of the quantum Fourier transform which creates an equal superposition of the states in the first register, we use the following operator:
\begin{equation}
U_{f_1} = \left(I^{\otimes m} - 2\ket{\bf{f_1}} \bra{\bf{f_1}}\right) 
\otimes I^{\otimes n},
\end{equation}
where $\bra{\bf{f_1}}=1/\mu[\kappa, 1,\dots, 1]$ with $\kappa$ being a coefficient and $\mu$ being a normalization constant.
The number of iterations can be minimized  by engineering the value of $\kappa$. 

\begin{figure}
\centering
\includegraphics[width = 3.2in]{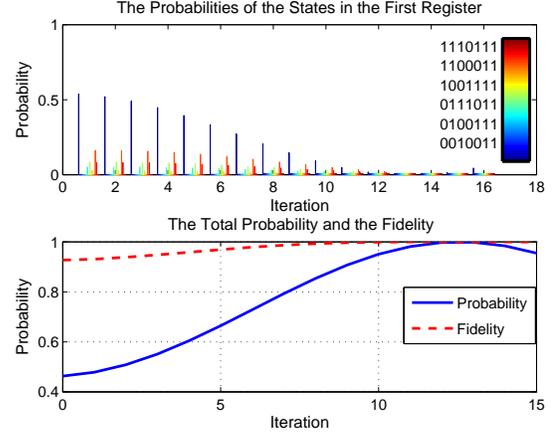}
\caption{The iterations of the amplitude amplification with the standard phase estimation algorithm for a random $16\times 16$ matrix with six non-zero eigenvalues.\label{figRandomwithqft}}
\end{figure}

\begin{figure}
\centering
\includegraphics[width = 3.2in]{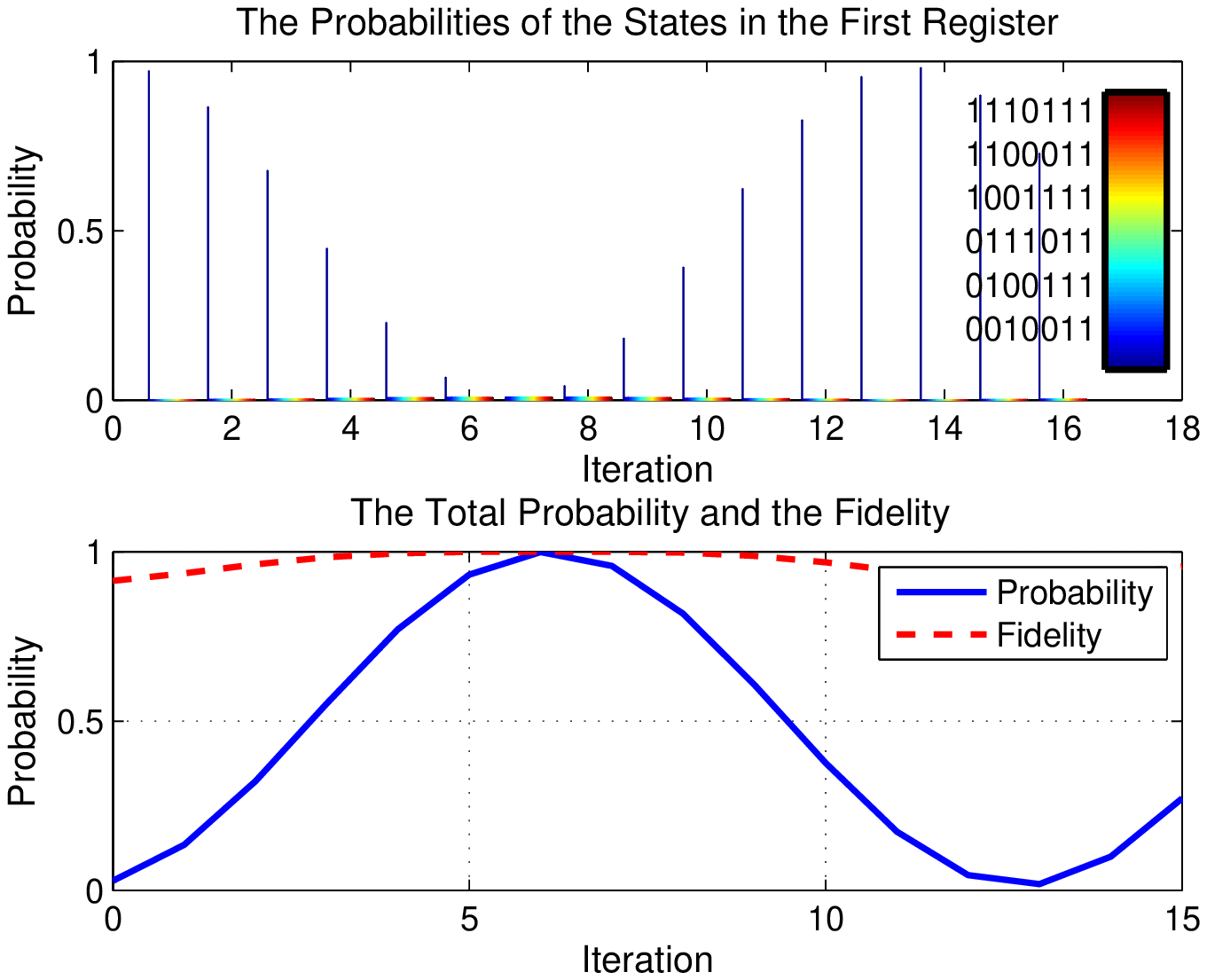}
\caption{The iterations of the amplitude amplification with the biased phase estimation algorithm for the same matrix as in Fig.\ref{figRandomwithqft}: in the construction of $U_{f_1}$, the value of $\kappa = 1$ and the initial success probability is around 0.028. \label{figBPEAwithRandom1}}
\end{figure}

\begin{figure}
\centering
\includegraphics[width = 3.2in]{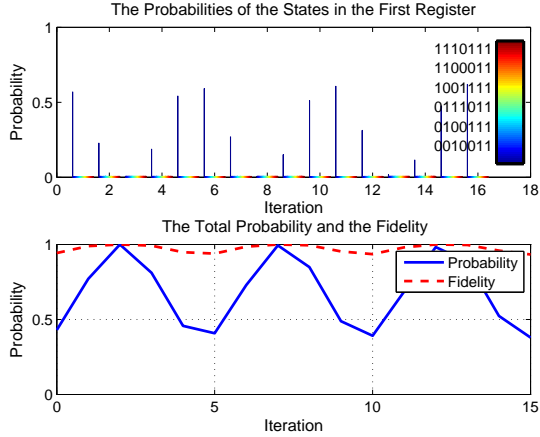}
\caption{The iterations of the amplitude amplification with the biased phase estimation algorithm for the same matrix as in Fig.\ref{figRandomwithqft}: in the construction of $U_{f_1}$, the value of $\kappa = 20$ and the initial success probability is around 0.43. \label{figBPEAwithRandom20}}
\end{figure}

Using $U_{f_1}$ in lieu of the quantum Fourier transform leads the algorithm drawn in Fig.\ref{figBPEA}, which produces a biased superposition of the eigenpairs in the output which speeds-up the amplitude amplification:
The amplification is done by inversing the marked item about the mean amplitude. As the amplitude of the marked item becomes closer to the mean amplitude, the amplification process gets slower (i.e., it requires more iterations). In fact, when $\kappa= \sqrt{M}$, $\kappa/\mu \approx (\sqrt{P_{\ket{0}}}+\sqrt{1-P_{\ket{0}}})/2.$: i.e. the mean amplitude. In this case,  no increment or decrement will be seen in the amplitude amplification.  
Therefore, making the initial probability further away from the mean probability, we can speed-up the amplitude amplification.
As an example, we show the iterations of the amplitude amplification with the standard phase estimation in Fig.\ref{figRandomwithqft} and with the biased phase estimation using $U_{f_1}$ with $\kappa=1$, and $\kappa = 20$, respectively in Fig.\ref{figBPEAwithRandom1} and 
Fig.\ref{figBPEAwithRandom20}:
As seen from Fig.\ref{figRandomwithqft}, the probability and the fidelity in PEA with QFT are maximized after 12 iterations. Here, the fidelity is defined as the overlap of the expected output $VV^T\ket{\mathbf{y}}$ with the actual output.
However, the maximum probability and the fidelity can be observed after 6 iterations  when we use $U_{f_1}$ with $\kappa=1$, which generates a very low initial success probability but approaches to the maximum faster.
Using  $U_{f_1}$ with $\kappa=20$ to increase the initial success probability also decreases the required number of iterations upto 2 iterations.

At the end of the phase estimation algorithm either the biased or the standard, we have 
${VV^T}\ket{\bf{y}}$, where a column of ${V}$ represents a principal eigenvector (eigenvector corresponding to a non-zero eigenvalue). 
$\bra{\bf{y}}{VV^T}\ket{\mathbf{y}}$ provides a similarity measure for  the data point ${\bf{y}}$ to the ${XX^T}$.
 $\bra{\mathbf{y}}{VV^T}\ket{\mathbf{y}}$ can be obtained either measuring the final state in a such basis in which $\ket{\mathbf{y}}$ is the first basis vector. Or one can apply a Householder transformation 
 ${I} - 2\ket{\mathbf{y}}\bra{\mathbf{y}}$ to the final state and then measure this state in the standard basis. 
In that case, the probability for measuring \ket{\mathbf{0}} yields the nearness (similarity) measure. 

Below, we summarize the steps of the biased phase estimation and then the amplitude amplification algorithms: 
\begin{tcolorbox}[enhanced,colback=lightgray!20!white,frame hidden,breakable,
  enlargepage flexible=\baselineskip
  ]
 As shown in Fig.\ref{figBPEA}, the steps of the biased phase estimation, $U_{BPEA}$, are as follows:
\begin{enumerate}
\item First, assume we have a mechanism to produce $U^{2^j}$ used in the phase estimation algorithm. Here, $U$ represents the circuit design for the data matrix. In the next section, we will discuss how to find this circuit by writing the data matrix as a sum of unitary matrices. 
\item Then, prepare the input state $\ket{\bf{0}}\ket{\bf{0}}$, where \ket{\bf{0}} represents the first vector in the standard basis. 
The phase estimation uses two registers: the phase and the system register. 
In our case, the final state of the system register becomes equivalent to the state obtained from $VV^T\ket{\bf{y}}$. 

\item To prepare  $\ket{\bf{y}}$ on the system register, apply an input preparation gate, $U_{input}$, to the system register.
\item Then, apply $U_{f_1} = I - 2\ket{\bf{f_1}}\bra{\bf{f_1}} \otimes I^{\otimes{n}}$ to put the phase register into a biased superposition.
\item Then, apply controlled $U^{2^j}$s to the system register.
\item Finally, apply $U_{f_1}^\dagger$ to get the eigenpairs on the output with some probability.
\end{enumerate}
\end{tcolorbox}

\begin{tcolorbox}[enhanced,colback=lightgray!20!white,frame hidden,breakable,
  enlargepage flexible=\baselineskip
  ]
The steps of the amplitude amplification process are as follows:
\begin{enumerate}
\item After applying $U_{BPEA}$ to the initial state  $\ket{\bf{0}}\ket{\bf{0}}$, apply the following iteration operator:
\begin{equation}
 Q = U_{BPEA}U_s U_{BPEA} U_{f_2}. 
\end{equation} 
\item Measure one of the qubits in the phase register: 
\begin{itemize}
\item if it is close to the equal superposition state, then stop the algorithm.
\item if it is not in the superposition state, then apply the iteration operator given above again.
\item Repeat this step until one of the qubits in the phase register is in the equal superposition state (Individual qubits present the same behavior as the whole state of the register. If one of  the qubits approaches to the equal superposition state, it generally indicates all qubit approaches to the equal superposition state.). 
\end{itemize}
\item Measure the second register in the basis in which $\ket{\bf{y}}$ is a basis vector. 
\end{enumerate}
\end{tcolorbox}

\section{Circuit Implementation and Computational Complexity}
\subsection{Circuit for $XX^T$ and Complexity Analysis}
Most of the Laplace matrices used in the clustering algorithm are generally very large but sparse. 
The sparse matrices can be simulated in polynomial time (polynomial in the number of qubits) on quantum computers(see e.g. Ref.\cite{Berry2007sparse,Andrew2011}).
When the matrix $H$ is dense (from now on we will use $H$ to represent $XX^T$ or $L$), the phase estimation algorithm requires $e^{iH}$. This exponential is generally approximated through the Trotter-Suzuki decomposition.
However, the order of the decomposition increases the complexity dramatically. 
Recently in Ref.\cite{Daskin2017SumofRankOne}, we have showed that one can use a matrix directly in the phase estimation by converting the matrix into the following form:
\begin{equation}
\widetilde{H} = I - i\frac{H}{k},
\end{equation}
where $k$ is a coefficient equal to $||H||_1\times 10$, which guarantees that the imaginary parts of the eigenvalues of $\widetilde{H}$ represents the eigenvalues of $H/k$ and are less than 0.1 so that $sin(\lambda_j/k)\approx \lambda_j/k$. Here, $\lambda_j$ is an eigenvalue of $H$.
When phase estimation algorithm is applied to $\widetilde{H}$, the obtained phase gives the value of $\lambda_j/k$.

The circuit implementation of $\widetilde{H}$ can be achieved by writing 
$\widetilde{H}$  as a sum of unitary matrices which can be directly mapped to quantum circuits.
This sum can be produced through the Householder transformation matrices formed by the data vectors. 
For the matrix $H = XX^T$, this can be done as follows:  \cite{Daskin2017SumofRankOne}:
\begin{equation}
H  =  -\frac{1}{2}\sum_{j=1}^L \left[ \left(I - 2\ket{\bf{x_j}}\bra{\bf{x_j}} \right) - I \right] 
\end{equation}
In the above, $H$ consists of $(L+1)$ number of Householder matrices each of which can be implemented by using $O(N)$ number of quantum gates. 
$H$ can be generated in an ancilla based circuit combining the circuits for each Householder matrix. The resulting circuit will require $O(LN)$ quantum operations.
In the phase estimation, if we use $m$ number of qubits in the first register, the complexity of the whole process becomes $O(2^mLN)$. 
In the case of $H$ written as a some of simple unitaries which can be represented by $O(logN)$ quantum gates,  
the complexity can be further reduced to $O(2^mLlog(N))$ number of quantum gates. 
\subsection{Comparison to Classical Computers}
Spectral clustering on classical computers requires some form of the eigendecomposition  of the matrix $H$ or at least the singular value decomposition of the matrix $X$. 
The classical complexity in general form can be bounded by $O(L^3 + LN)$, where $O(L^3)$ is considered complexity for the eigendecomposition of the matrix and $LN$ for processing the matrix $X$ at least once.
As a result, if $H$ can be written in terms of simple unitaries, then quantum computers is likely to provide a dramatic improvement in the computational complexity of spectral clustering.
However, on the other cases, the complexity of the clustering is not much different than the classical complexity even though improvements specific to some applications are still possible.
Note that since the amplitude amplification is run only a few times, 
it does not change the bound of the complexity.
However, we need to run the whole process a few times with different $\ket{\bf{y}}$ indicator vectors so as to find the maximum similarity and determine the right clustering.
Therefore, there is an additional coefficient $c$ indicating the number of trials in the complexity: i.e. $O(c2^mLN)$. 
Since the coefficient $c$  also exists in the classical complexity, it does not change the argument in the comparison of the complexities.

 \subsection{Measurement of the Output}

If only tensor product of the eigenvectors of Pauli operators $\{I,\sigma_x,\sigma_y,\sigma_z \}$ are used, 
the number of nonzero elements  can be either one of the followings: $N, N/2, \dots 1$ to produce 
 a disentangled quantum state \ket{\mathbf{y}}, which can be efficiently implemented. 
We can also use an operator similar to $Y = \sum_i \sigma_x^i$, where $i$ indicates a $\sigma_x$ gate on the ith qubit of the system, to approximate the best clustering (This operator is also used in Ref.\cite{Farhi2014quantum} in the description of a quantum optimization algorithm).
In this case, the initial state is taken as an equal superposition state, then multiplied with $e^{iY}$. 
In the end, the operator  $e^{iY}$ is applied again. 
The measurement outcome yields the index of the column of $Y$ which can be considered as an approximate clustering solution.
Since the circuit implementation of  $e^{iY}$ requires only $n$ single $\sigma_x$-gates and the solution is obtained only from one trial, the total complexity becomes $O(2^mLN)$.

As mentioned before, the other measurements can be approximated through the mapping to the standard basis by applying the Householder transformation  ${I} - 2\ket{\mathbf{y}}\bra{\mathbf{y}}$ to the system register. In that case, the probability to see the system register in \ket{\bf{0}} represents the similarity measure.

\section{Conclusion}
Here, we have described how to do spectral clustering on quantum computers by employing the famous quantum phase estimation  and amplitude amplification algorithms.
We have shown the implementation steps and analyzed the complexity of the whole process.
In addition, we have shown that the required number of iterations in the amplitude amplification process can be dramatically decreased by  using a biased operator instead of the quantum Fourier transform in the phase estimation algorithm.

\bibliographystyle{IEEEtran}


\end{document}